%% file: ANA-HIGG-2022-20-PAPER.tex
\documentclass[PAPER, cernpreprint, atlasdraft=false, texlive=2023, USenglish, texmf, orcidlogo]{atlasdoc}

\usepackage{atlaspackage}
\usepackage{atlasbiblatex}
 
\usepackage{atlasphysics}

\graphicspath{{logos/}{figures/}}
 
\usepackage{ANA-HIGG-2022-20-PAPER-defs}

\AtlasTitle{Combined measurement of the Higgs boson mass from the $H\to\gamma\gamma$ and $H\to ZZ^{*} \to 4\ell$ decay channels with the ATLAS detector using $\sqrt{s}$ = 7, 8 and 13 TeV $pp$ collision data}

\AtlasAbstract{
A measurement of the mass of the Higgs boson combining the $H\to ZZ^{*} \to 4\ell$ and $H\to\gamma\gamma$ decay channels is presented.  The result is based on 140 fb$^{-1}$ of proton-proton collision data collected by the ATLAS detector during LHC Run 2 at a centre-of-mass energy of 13 TeV combined with the Run 1 ATLAS mass measurement, performed at centre-of-mass energies of 7 and 8 TeV, yielding a Higgs boson mass of  125.11  $\pm$ 0.09 (stat.) $\pm$ 0.06 (syst.) = 125.11 $\pm$ 0.11 GeV. This corresponds to a 0.09\% precision achieved on this fundamental parameter of the Standard Model of particle physics.}

\AtlasRefCode{HIGG-2022-20}
 
\PreprintIdNumber{CERN-EP-2023-156}

\AtlasJournal{Phys.~Rev.~Lett.}
\AtlasJournalRef{Phys. Rev. Lett. 131 (2023) 251802}
\AtlasDOI{10.1103/PhysRevLett.131.251802}

\hypersetup{pdftitle={ATLAS document},pdfauthor={The ATLAS Collaboration}}
 
\begin{document}
 
\maketitle

 
The discovery of the Higgs boson in proton-proton ($pp$) collisions at the CERN LHC by the ATLAS and CMS Collaborations \cite{ATLAS:2012yve,CMS:2012qbp} with data collected at $\sqrt{s}$ = 7 \TeV\ and 8 \TeV\ (Run 1) was a major step towards understanding the electroweak symmetry breaking mechanism.
Gauge theories require in fact that gauge bosons be massless, in apparent contradiction with observations. In this context, the seminal work of Brout, Englert~\cite{Englert:1964et}, Higgs~\cite{Higgs:1964ia,Higgs:1964pj,Higgs:1966ev} and Guralnik, Hagen and Kibble~\cite{Guralnik:1964eu, Kibble:1967sv}, has provided a consistent mechanism for the generation of gauge boson masses. The Glashow-Weinberg-Salam theory extended this mechanism proposing a theory of the electroweak interactions~\cite{Glashow:1961tr,Salam:1968rm,Weinberg:1967tq}, introducing a doublet of complex scalar fields, which couples also to fermions, providing them with a mass that would otherwise be absent. This forms a major component of the Standard Model (SM) of particle physics. A salient prediction of the SM is the presence of a Higgs boson, whose mass is not predicted by the theory and needs to be estimated experimentally. Since the Higgs boson discovery, thanks to the luminosity accumulated at the LHC between 2015 and 2018 (Run 2) and the increased center-of-mass energy at $\sqrt{s}$ = 13 \TeV, the focus has shifted to the precise measurements of Higgs boson properties~\cite{ATLAS:2022vkf, CMS:2022dwd}. The couplings of the Higgs boson to other elementary particles are predicted in the SM once the Higgs boson mass \mh\ is known. This motivates its precise measurement through decay channels that can be fully reconstructed and with the best mass resolution.
 
The \Hyy\ and \Hllll\ decays are the most suitable processes to measure \mh\ at the LHC due to their excellent mass resolution, which produce a clear mass peak above a continuum background~\cite{ATLAS:2012yve,CMS:2012qbp}. The Higgs boson mass \mh\ was measured by ATLAS and CMS in the \Hyy\ and \Hllll\ channels using the full Run 1 dataset, and all measurements by the two experiments were combined resulting in a \mh\ value of 125.09 $\pm$ 0.24 \GeV\ \cite{ATLAS:2015yey}. More recently, the CMS Collaboration has measured \mh\ in the same decay channels using 35.9 \ifb\ of 13 TeV $pp$ Run 2 collision data. The combination of the two CMS Run 2 measurements with their Run 1 results yielded a \mh\ value of 125.38 $\pm$ 0.14 \GeV\ \cite{CMS:2020xrn}.
This letter presents a measurement of \mh\ combining the \Hyy\ \cite{HIGG-2019-16-PAPER} and \Hllll\ \cite{HIGG-2020-07} decay channels. The result is based on 140 fb$^{-1}$ of proton-proton collision data collected by the ATLAS detector~\cite{PERF-2007-01} during the LHC Run 2 at a centre-of-mass energy of 13 TeV, and updates and supersedes that based on the same final states and a partial Run 2 dataset corresponding to an integrated luminosity of 36.1 fb$^{-1}$ \cite{HIGG-2016-33}.
An extensive software suite~\cite{ATL-SOFT-PUB-2021-001} is used in the reconstruction and analysis of real and simulated data, in detector operations, and in the trigger and data acquisition systems of the experiment.
The combined measurements profit from the increased dataset, and from significantly improved calibrations of the electron and photon energy \cite{HIGG-2019-16-PAPER,EGAM-2021-02-PAPER} and of the muon momentum \cite{HIGG-2020-07,ATLAS:2022jjr}.


 
The mass measurement reported in this letter is performed  using  the profile likelihood ratio  \cite{ATL-PHYS-PUB-2011-011,Cowan:2010js} defined as $\Lambda(m_{H})=\mathcal{L}(m_{H},\hat{\hat{\pmb{\theta}}}(m_{H}))/\mathcal{L}(\hat{m}_{H},\hat{\pmb{\theta}})$. $\hat{m}_{H}$ and $\hat{\pmb{\theta}}$ are the values of the parameter of interest and nuisance parameters (NP) that maximize the likelihood $\mathcal{L}(m_{H}, \pmb{\theta})$, while $\hat{\hat{\pmb{\theta}}}(m_{H})$ corresponds to the values of the NP that maximize the likelihood for a given value of \mh. Systematic uncertainties are modelled by constrained NP, while the signal and background normalizations in the various channels entering the fit are treated as free parameters.
The confidence intervals are obtained assuming the asymptotic distribution of the $-2\ln{\Lambda(m_H)}$ test statistic \cite{Cowan:2010js}.
The statistical uncertainty on \mh\ is estimated by fixing all the NP that are associated with systematic uncertainties to their best-fit values and leaving all the remaining parameters unconstrained.
The total systematic uncertainty, whose squared value is evaluated as the difference between the squares of the total uncertainty and the statistical uncertainty, can be decomposed into categories representing distinct sources of uncertainty by setting all relevant subsets of NP to their best-fit values.


 
The full description of the Run 2 mass measurement in the \Hyy\ channel is given in Ref.~\cite{HIGG-2019-16-PAPER}. A description of the key aspects of this measurement is summarized here.
The \Hyy\ decay is reconstructed by requiring two energetic photons fulfilling strict identification and isolation criteria. The invariant mass \myy\ distribution of the selected photon pairs exhibits a peak near \mH, arising from resonant Higgs boson decays, over a smoothly falling distribution from background processes mainly due to non-resonant diphoton production. The value of \mh\ is determined from the position of the peak in data through a profile-likelihood fit to the $\myy$ distribution. Simulated signal and background event samples are used to optimize the analysis criteria, to choose the signal and background $\myy$ models used in the fit, and to estimate some of the systematic uncertainties on \mH. To increase the sensitivity of the measurement, the selected events are classified into 14 mutually exclusive categories with different diphoton invariant mass resolutions and signal-to-background ratios, which are analyzed simultaneously. The normalization factor for each category is independent and fitted to the data. The $\myy$ resolution ranges from about 1.1 \GeV\ to 2.0 \GeV, depending on the category. The signal model consists of a double-sided Crystal Ball probability density function \cite{Oreglia:1980cs}, with the mean and standard deviation of its Gaussian core parameterized as a function of \mh\ in each category using simulated signal events generated at different \mh\ hypotheses.
Compared with the mass result reported in Ref.~\cite{HIGG-2016-33}, the \Hyy\ mass measurement used in this combination and reported in Ref.~\cite{HIGG-2019-16-PAPER} profits from an increased data sample, a new photon reconstruction algorithm with better energy resolution~\cite{EGAM-2018-01}, an improved estimation of the photon energy scale with significantly reduced uncertainties~\cite{EGAM-2021-02-PAPER}, and an optimized event classification strategy.
Uncertainties for photons converting into electron-positron pairs before reaching the electromagnetic calorimeter, which are experimentally similar to electrons, are only moderately improved by the updated calibrations at energies typically observed in the \Hyy\ decay (e.g. $\et\sim 60~\GeV$). For unconverted photons, the energy calibration is improved by typically 30\% in the central part of the calorimeter ($|\eta| < 1.37$), and up to a factor 2 in the endcap region ($1.51 \leq |\eta| < 2.37$).
The reduction of the uncertainties on the photon energy scale arises from an improved understanding of the difference in data and simulation of the inputs to the photon energy scale regression, and of the introduction of transverse energy (\ET) dependent \emph{in-situ} scales derived from $Z\to e^+e^-$ events, that reduce the calibration extrapolation uncertainties from the $Z$ boson mass to the Higgs mass and from electrons to photons \cite{EGAM-2021-02-PAPER}.
The measured mass of the Higgs boson in the \Hyy\ final state using the full Run 2 dataset is \mhyyrunB\ \cite{HIGG-2019-16-PAPER}.
The dominant sources of systematic uncertainties on the measurement are associated to the $Z\to e^+e^-$ \emph{in-situ} scale (59 \MeV), the residual \ET-dependent electron energy scale calibration (44 \MeV), and the calibration extrapolation from electrons to photons (30 \MeV) \cite{HIGG-2019-16-PAPER}.
The effect of the interference between the $\Hyy$ signal and the $\gamma\gamma$ continuous background \cite{Dixon:2013haa} is evaluated to have an impact on the determination of \mH\ of approximately 26 \MeV. The full effect is accounted as a systematic uncertainty on the quoted result, and no shift of the mass value is applied.
A combination with the measurement of \mh\ using the Run 1 dataset~\cite{ATLAS:2015yey}, \mhyyrunA, is performed. In this combination, only the \ET-independent component of the uncertainty associated to the \emph{in-situ} scale derived from $Z\to e^+e^-$ events, the resolution uncertainties, and the theoretical uncertainties related to the various Higgs production modes are considered as correlated  between Run 1 and Run 2. The combined measurement of \mh\ using  Run 1 and Run 2 datasets in the \Hyy\ channels is \mhyy.


 
The full description of the Run 2 mass measurement in the \Hllll\ channel is given in Ref.~\cite{HIGG-2020-07}. A description of the key aspects of this measurement is summarized here.
The \Hllll\ decay is reconstructed by requiring two  pairs of same-flavor opposite-sign isolated leptons ($\ell$=$e$,$\mu$) in the final state. The pair with the invariant mass closer to that of the Z boson mass is defined as the leading dilepton pair, while the remaining one is referred to as the subleading dilepton pair. The selected quadruplets are separated into four sub-channels according to the flavor of the leading and subleading dilepton pairs ($4\mu$, $2e2\mu$, $2\mu2e$, $4e$). A neural-network-based classifier  is employed to discriminate between the Higgs boson signal and the dominant $ZZ^{*}\to 4\ell$ background. The \mh\ measurement is performed with a simultaneous unbinned maximum likelihood fit of the reconstructed invariant mass of  the four leptons system, $m_{4\ell}$, in the four sub-channels. The  $m_{4\ell}$ resolution ranges from about 1.5 GeV ($4\mu$ and $2e2\mu$ sub-channels) to about 2.1 GeV  ($2\mu 2e$ and $4e$ sub-channels). The signal model consists of a double-sided Crystal Ball  probability density function, with the mean of its Gaussian core parameterized as a function of \mh\ and the standard deviation expressed as a function of the predicted event-level resolution. The signal and background normalization for each of the four sub-channels are free parameters in the fit.
Compared with the measurement reported in Ref.~\cite{HIGG-2016-33}, the \Hllll\  used in this combination and reported in Ref.~\cite{HIGG-2020-07} profits from an increased data sample, a new high-precision muon momentum calibration~\cite{ATLAS:2022jjr}, the neural-network-based classifier for the signal versus background discrimination and the inclusion of the event-by-event invariant mass resolution in the analytical model used to fit the collision data.
The measured mass of the Higgs boson in the \Hllll\ final state using the full Run 2 dataset is \mhzzrunB. The dominant sources  of systematic uncertainty on the measurement are the uncertainties in the muon momentum scale, resolution and sagitta bias correction  ($28$ MeV) and the electron energy scale ~\cite{EGAM-2018-01} ($19$ MeV). A combination with the measurement of \mh\ using the Run 1 dataset~\cite{ATLAS:2015yey}, \mhzzrunA\ has been performed. In this combination, only the uncertainties on the electron calibration were considered correlated, while the muon calibration systematic uncertainty is uncorrelated between the two measurements due to improved and independent techniques in the muon momentum scale calibration. The combined measurement of \mh\ performed with Run 1 and Run 2 datasets in the \Hllll\ channel is \mhzz.


 
The combined mass measurement in the \Hyy\ channel \cite{HIGG-2019-16-PAPER} is similarly affected by the statistical uncertainty (110 \MeV) and the systematic uncertainty (90 \MeV), mainly associated to the photon energy scale calibration. In contrast, the combined mass measurement in the \Hllll\ channel \cite{HIGG-2020-07} is primarily dominated by the statistical uncertainty (170 \MeV), while the systematic uncertainty, mainly related the muon momentum scale calibration, has a minor impact (30 \MeV) on the measurement. The differences between the two channels can be traced to the distinct decay branching ratios, final state reconstruction efficiencies, background levels, and the resulting signal-to-background ratios in the two channels. A detailed comparison of the two channels, qualitatively similar to those presented here, is given in Ref.~\cite{HIGG-2013-12}.


 
\begin{figure}
\centering
\includegraphics[width=.48\textwidth]{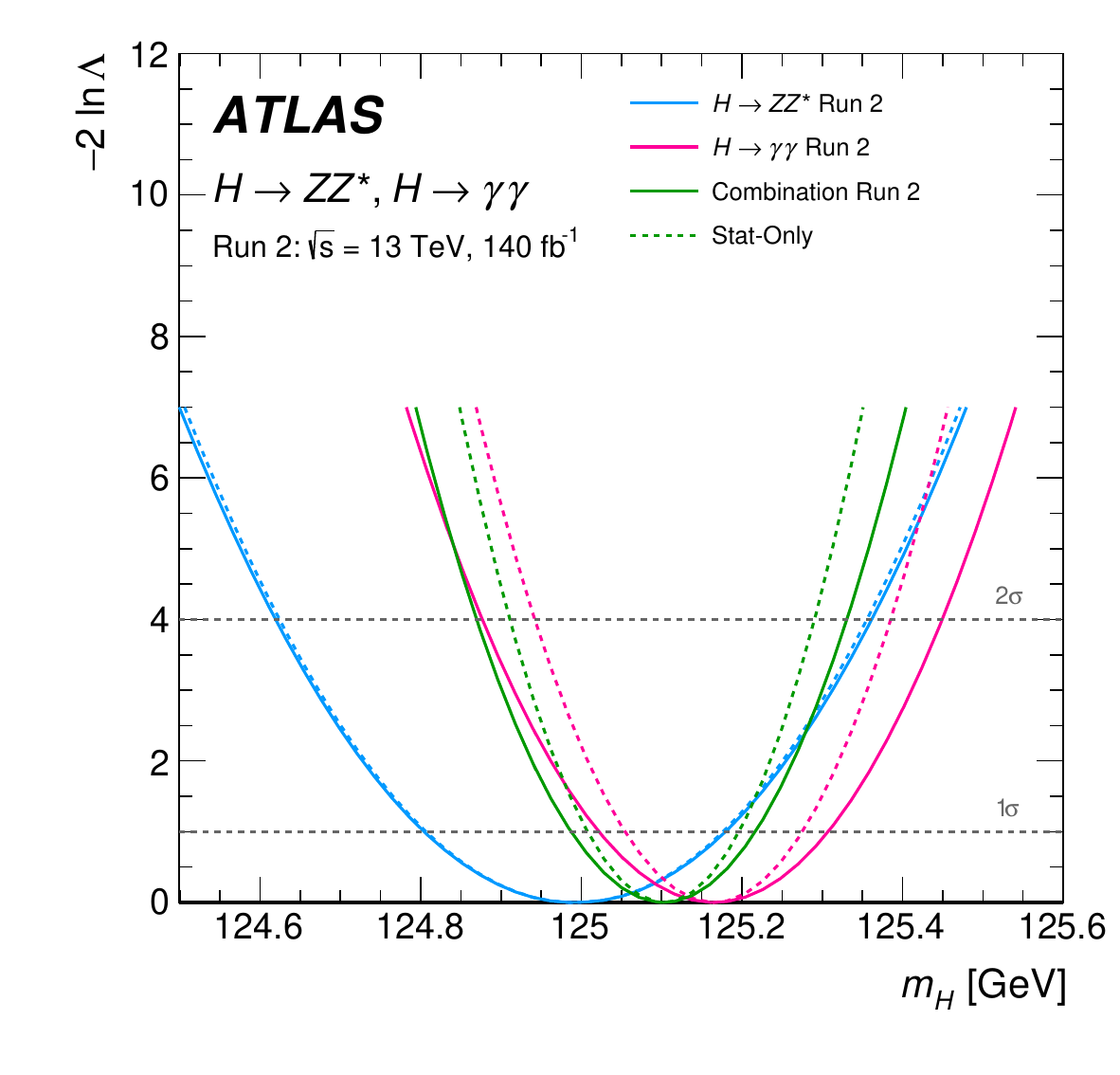}
\includegraphics[width=.48\textwidth]{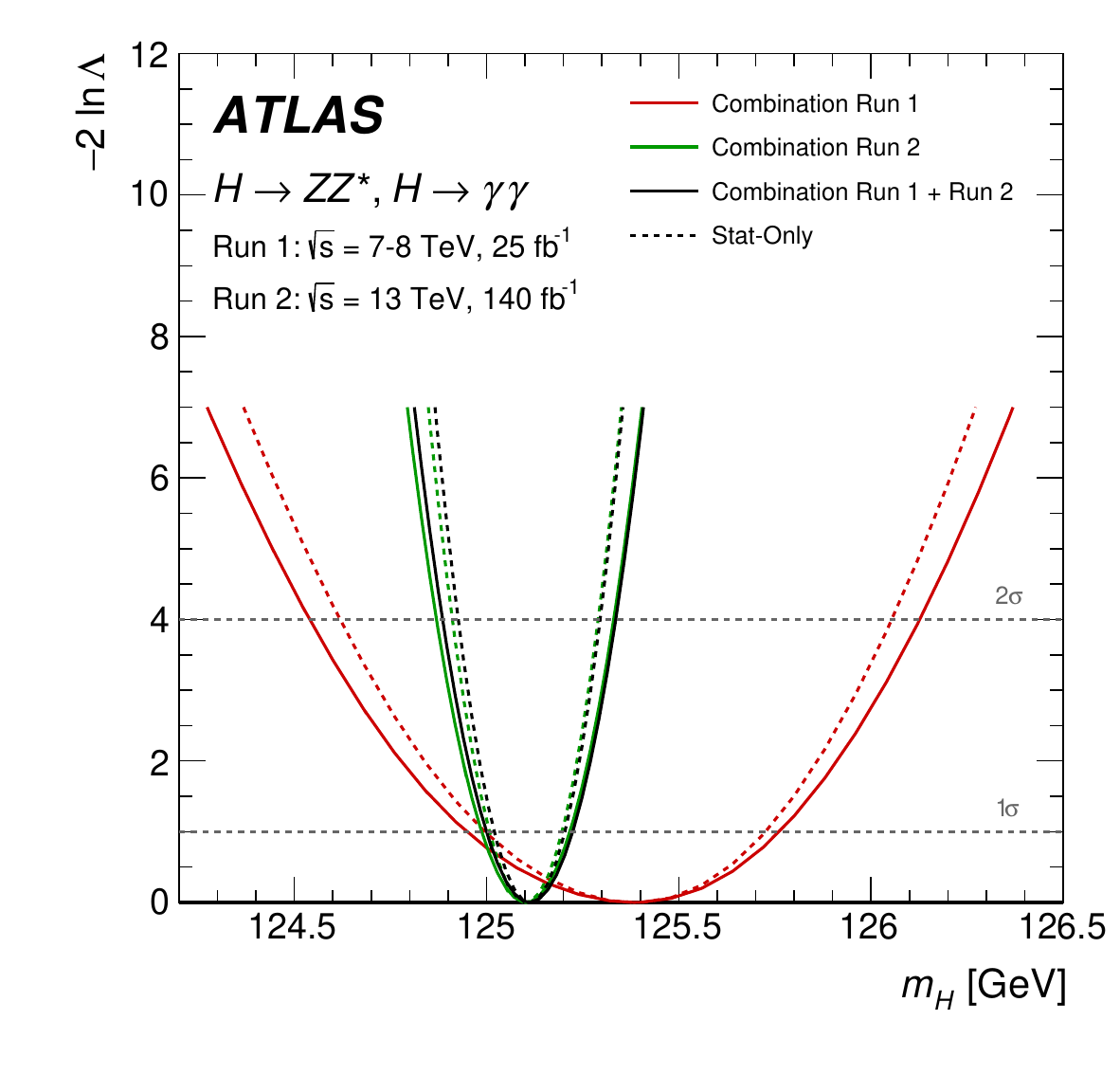}
\caption{Value of $-2\ln\Lambda$ as a function of \mh\ for (left) \Hyy, \Hllll\ channels and their combination (magenta, cyan and green, respectively) using Run 2 data only and for (right) Run 1, Run 2 and their combination (red, green and black, respectively). The dashed lines show the statistical component of the uncertainty. The 1$\sigma$ (2$\sigma$) confidence interval is indicated by the intersections of the horizontal line at 1 (4) with the log-likelihood curves.}
\label{fig:nll}
\end{figure}
 
\begin{figure}
\centering
\includegraphics[width=0.8\textwidth]{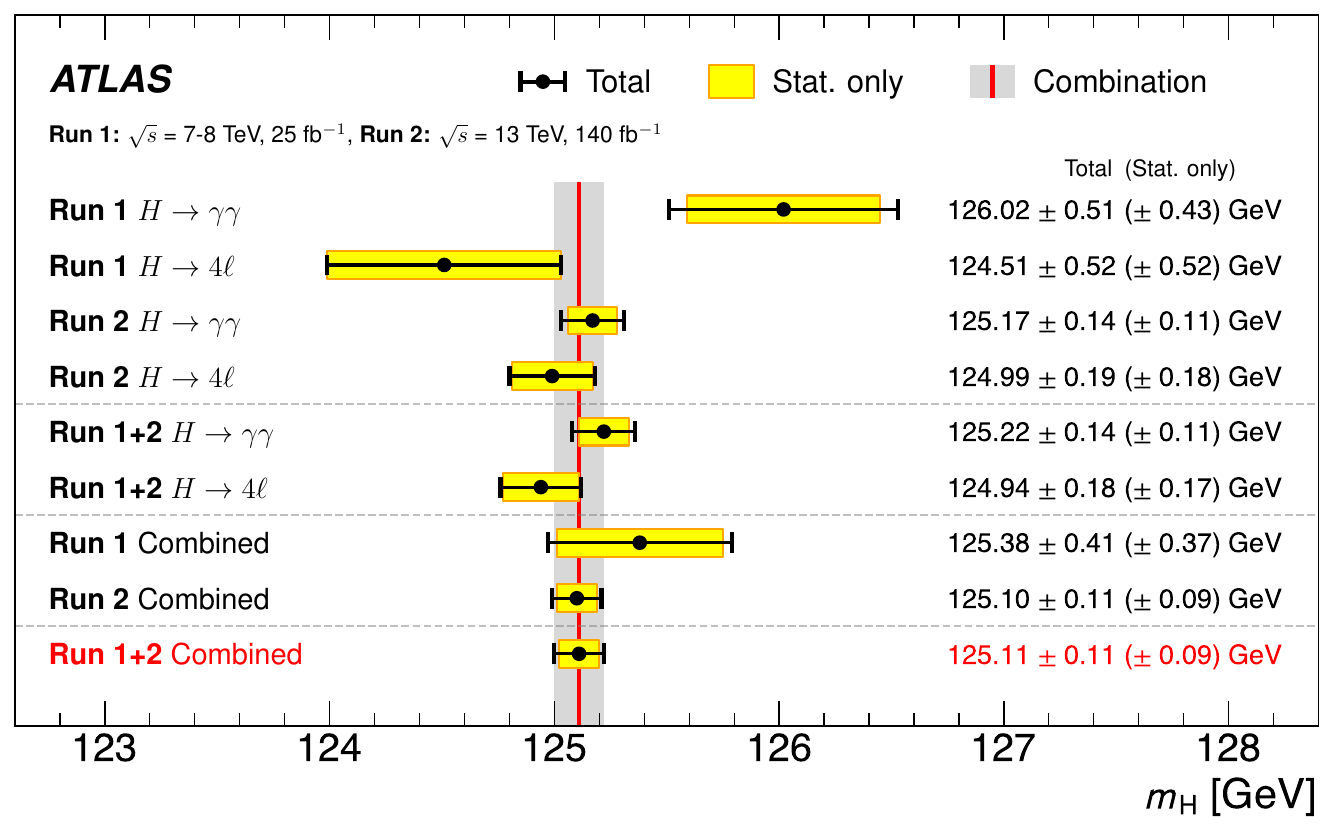}
\caption{Summary of \mh\ measurements from the individual \Hyy\ and \Hllll\ channels and their combination presented in this letter. The uncertainty bar on each point corresponds to  the total uncertainty; the horizontal shaded bands represent the statistical component of the uncertainties; the vertical red line and gray band represent the combined result presented in this letter with its total uncertainty.}
\label{fig:summary}
\end{figure}
 
In the \Hyy\ and \Hllll\ channels combination, the correlations between systematic uncertainties in the two measurements are accounted for in the profile likelihood function by using the same constraint for each of the correlated NP. All potential correlations between measurements and data-taking periods are thoroughly examined.
Due to substantial variations in the calibration of electrons, photons, and muons, most correlations are small.
If applicable, these correlations are incorporated following the approach that yields the most conservative result.
In the combinations of the Run 1 and Run 2 measurements of the \Hyy\ and \Hllll\ individual channels, the correlation of the experimental systematic uncertainties follows what was done in  Refs.~\cite{HIGG-2019-16-PAPER} and \cite{HIGG-2020-07}, respectively.
The correlation scheme between the Run 1 \Hyy\ and \Hllll\ measurements is unchanged relative to the published Run 1 combination~\cite{ATLAS:2015yey}.
The choice of correlation model between the Run 2 \Hyy\ and \Hllll\ measurements reflects the improvements in the photon calibration adopted by the \Hyy\ analysis not being mirrored in the calibration of the electrons used in the \Hllll\ analysis; only the electron and photon resolution systematic uncertainties and those associated with the \ET-independent component of the electron and photon \emph{in-situ} energy scale are considered as correlated.
Other sources of systematic uncertainties correlated between the two channels are the theory uncertainties on the prediction of the various Higgs production modes, the modelling of additional (\textit{pile-up}) $pp$ collisions, and the uncertainty on the integrated luminosity. 
The choice of correlation model is also tested by using different approaches (e.g. correlating the muon calibration systematic uncertainties in Run 1 and Run 2, correlating all sources of photon and electron calibration systematic uncertainties between the \Hyy\ and \Hllll\ channels) and is shown to have negligible impact on the result.
Signal yield normalizations  are treated as independent free parameters in the fit to minimize model-dependent assumptions in the measurement of \mh.
 
The combined value measured using Run 2 data is \mhrunBshort.
The uncertainty is compatible with the expected error assuming a SM Higgs boson mass of 125 \GeV.
The statistical component of the uncertainty is $\pm$ 0.09 \GeV. The corresponding profile likelihood, for the two channels and for their combination, is shown in Figure~\ref{fig:nll} (left) as a function of \mH.
If the small interference predicted by the SM between the Higgs boson and the non-resonant di-photon background was considered for the \Hyy\ signal parameterization, the \mh\ value measured by the combination would increase by 15 \MeV.
This result is in good agreement with the ATLAS + CMS Run 1 measurement \cite{HIGG-2016-33}, $m_H = 125.09 \pm 0.24$ \GeV.
The contributions of the main sources of systematic uncertainty to the combined measurement, using ATLAS Run 2 data, are summarized in Table~\ref{tab:systematics}. The values differ from those reported in Refs.~\protect\cite{HIGG-2019-16-PAPER} and \protect\cite{HIGG-2020-07} because of the relative impact of the \Hyy\ and \Hllll\ results in the combination. The \ET-independent component of the electron and photon \emph{in-situ} energy scale (``$e/\gamma$ \ET-independent $Z\to ee$ calibration'' in Table~\ref{tab:systematics}) is among the few uncertainties correlated between the \Hyy\ and \Hllll\ measurements, and impacts the former measurement by 59 \MeV\ \cite{HIGG-2019-16-PAPER}, and the latter by 19 \MeV\ \cite{HIGG-2020-07}.
The combined measurement from the ATLAS Run 1 and Run 2 results is \mhrunABshort. The statistical component of the uncertainty is $\pm$ 0.09 \GeV. The four combined measurements are compatible with a $p$-value of $18\%$. Figure~\ref{fig:nll} (right) shows the corresponding profile likelihoods, separately for the ATLAS Run 1 and Run 2 datasets, as well as for their combination, as a function of \mH.
The contributions of the main sources of systematic uncertainty to the combination of Run 1 and Run 2 data are nearly identical to those presented in Table~\ref{tab:systematics}.
Figure~\ref{fig:summary} presents a summary of the \mh\ measurements from the individual \Hyy\ and \Hllll\ channels and their combinations discussed in this letter.

\begin{table}
\centering
\begin{tabular}{lc}
\hline \hline
Source                                           & Systematic uncertainty on \mH\ [\ensuremath{\text{M\electronvolt}}] \\
\hline
$e/\gamma$ \ET-independent $Z\to ee$ calibration & 44                                    \\
$e/\gamma$ \ET-dependent electron energy scale   & 28                                    \\
\Hyy\ interference bias                          & 17                                    \\
$e/\gamma$ photon lateral shower shape           & 16                                    \\
$e/\gamma$ photon conversion reconstruction      & 15                                    \\
$e/\gamma$ energy resolution                     & 11                                    \\
\Hyy\ background modelling                       & 10                                    \\
Muon momentum scale                              & 8                                     \\
All other systematic uncertainties               & 7                                     \\
\hline \hline
\end{tabular}
\caption{Impact of the main sources of systematic uncertainty on the \mH\ measurement from the combination of the \Hllll\ and \Hyy\ final states using Run 2 data.
The systematic uncertainties associated with the combination of Run 1 and Run 2 data are nearly identical. The sum in quadrature of the individual contributions is not expected to reproduce the total systematic uncertainty due to the different methodologies employed to derive them.}
\label{tab:systematics}
\end{table}


 
In conclusion, the Higgs boson mass \mh\ is measured using Run 2 collision data at 13 \TeV\ yielding:
\begin{equation*}
\mhrunB ,
\end{equation*}
which is a significant improvement with respect to that reported in Ref.~\cite{HIGG-2016-33}. The systematic uncertainty affecting the \Hyy\ measurement is reduced by a factor of about three thanks to a novel and improved approach to the photon energy calibration. This is comparable with the factor of about two associated with  the increase in the data statistics. The systematic uncertainty on the muon momentum calibration decreases by about 50\% relative to Ref.~\cite{HIGG-2016-33}.
Combining the Run 2 result with the \mh\ measurements performed in Run 1 at 7 and 8 \TeV\ , the combined result is:
\begin{equation*}
\mhrunAB .
\end{equation*}
This result currently represents the most precise measurement of the Higgs boson mass, reaching a 0.09\% precision on this fundamental quantity.



We thank CERN for the very successful operation of the LHC, as well as the
support staff from our institutions without whom ATLAS could not be
operated efficiently.
 
We acknowledge the support of
ANPCyT, Argentina;
YerPhI, Armenia;
ARC, Australia;
BMWFW and FWF, Austria;
ANAS, Azerbaijan;
CNPq and FAPESP, Brazil;
NSERC, NRC and CFI, Canada;
CERN;
ANID, Chile;
CAS, MOST and NSFC, China;
Minciencias, Colombia;
MEYS CR, Czech Republic;
DNRF and DNSRC, Denmark;
IN2P3-CNRS and CEA-DRF/IRFU, France;
SRNSFG, Georgia;
BMBF, HGF and MPG, Germany;
GSRI, Greece;
RGC and Hong Kong SAR, China;
ISF and Benoziyo Center, Israel;
INFN, Italy;
MEXT and JSPS, Japan;
CNRST, Morocco;
NWO, Netherlands;
RCN, Norway;
MEiN, Poland;
FCT, Portugal;
MNE/IFA, Romania;
MESTD, Serbia;
MSSR, Slovakia;
ARRS and MIZ\v{S}, Slovenia;
DSI/NRF, South Africa;
MICINN, Spain;
SRC and Wallenberg Foundation, Sweden;
SERI, SNSF and Cantons of Bern and Geneva, Switzerland;
MOST, Taiwan;
TENMAK, T\"urkiye;
STFC, United Kingdom;
DOE and NSF, United States of America.
In addition, individual groups and members have received support from
BCKDF, CANARIE, Compute Canada and CRC, Canada;
PRIMUS 21/SCI/017 and UNCE SCI/013, Czech Republic;
COST, ERC, ERDF, Horizon 2020 and Marie Sk{\l}odowska-Curie Actions, European Union;
Investissements d'Avenir Labex, Investissements d'Avenir Idex and ANR, France;
DFG and AvH Foundation, Germany;
Herakleitos, Thales and Aristeia programmes co-financed by EU-ESF and the Greek NSRF, Greece;
BSF-NSF and MINERVA, Israel;
Norwegian Financial Mechanism 2014-2021, Norway;
NCN and NAWA, Poland;
La Caixa Banking Foundation, CERCA Programme Generalitat de Catalunya and PROMETEO and GenT Programmes Generalitat Valenciana, Spain;
G\"{o}ran Gustafssons Stiftelse, Sweden;
The Royal Society and Leverhulme Trust, United Kingdom.
 
The crucial computing support from all WLCG partners is acknowledged gratefully, in particular from CERN, the ATLAS Tier-1 facilities at TRIUMF (Canada), NDGF (Denmark, Norway, Sweden), CC-IN2P3 (France), KIT/GridKA (Germany), INFN-CNAF (Italy), NL-T1 (Netherlands), PIC (Spain), ASGC (Taiwan), RAL (UK) and BNL (USA), the Tier-2 facilities worldwide and large non-WLCG resource providers. Major contributors of computing resources are listed in Ref.~\cite{ATL-SOFT-PUB-2023-001}.


\printbibliography
 
\clearpage
\input{atlas_authlist}

\clearpage

\end{document}

%% file: atlas_authlist.tex
 
\begin{flushleft}
\hypersetup{urlcolor=black}
{\Large The ATLAS Collaboration}

\bigskip

\AtlasOrcid[0000-0002-6665-4934]{G.~Aad}$^\textrm{\scriptsize 102}$,
\AtlasOrcid[0000-0002-5888-2734]{B.~Abbott}$^\textrm{\scriptsize 120}$,
\AtlasOrcid[0000-0002-1002-1652]{K.~Abeling}$^\textrm{\scriptsize 55}$,
\AtlasOrcid[0000-0001-5763-2760]{N.J.~Abicht}$^\textrm{\scriptsize 49}$,
\AtlasOrcid[0000-0002-8496-9294]{S.H.~Abidi}$^\textrm{\scriptsize 29}$,
\AtlasOrcid[0000-0002-9987-2292]{A.~Aboulhorma}$^\textrm{\scriptsize 35e}$,
\AtlasOrcid[0000-0001-5329-6640]{H.~Abramowicz}$^\textrm{\scriptsize 151}$,
\AtlasOrcid[0000-0002-1599-2896]{H.~Abreu}$^\textrm{\scriptsize 150}$,
\AtlasOrcid[0000-0003-0403-3697]{Y.~Abulaiti}$^\textrm{\scriptsize 117}$,
\AtlasOrcid[0000-0002-8588-9157]{B.S.~Acharya}$^\textrm{\scriptsize 69a,69b,q}$
% ...
\bigskip
\\

% ...
$^{29}$Physics Department, Brookhaven National Laboratory, Upton NY; United States of America.\\
% ...
$^{35}$$^{(a)}$Facult\'e des Sciences Ain Chock, R\'eseau Universitaire de Physique des Hautes Energies - Universit\'e Hassan II, Casablanca;$^{(b)}$Facult\'{e} des Sciences, Universit\'{e} Ibn-Tofail, K\'{e}nitra;$^{(c)}$Facult\'e des Sciences Semlalia, Universit\'e Cadi Ayyad, LPHEA-Marrakech;$^{(d)}$LPMR, Facult\'e des Sciences, Universit\'e Mohamed Premier, Oujda;$^{(e)}$Facult\'e des sciences, Universit\'e Mohammed V, Rabat;$^{(f)}$Institute of Applied Physics, Mohammed VI Polytechnic University, Ben Guerir; Morocco.\\
% ...
$^{49}$Fakult\"{a}t Physik , Technische Universit{\"a}t Dortmund, Dortmund; Germany.\\
% ...
$^{55}$II. Physikalisches Institut, Georg-August-Universit\"{a}t G\"ottingen, G\"ottingen; Germany.\\
% ...
$^{69}$$^{(a)}$INFN Gruppo Collegato di Udine, Sezione di Trieste, Udine;$^{(b)}$ICTP, Trieste;$^{(c)}$Dipartimento Politecnico di Ingegneria e Architettura, Universit\`a di Udine, Udine; Italy.\\
% ...
$^{102}$CPPM, Aix-Marseille Universit\'e, CNRS/IN2P3, Marseille; France.\\
% ...
$^{117}$Department of Physics, New York University, New York NY; United States of America.\\
% ...
$^{120}$Homer L. Dodge Department of Physics and Astronomy, University of Oklahoma, Norman OK; United States of America.\\
% ...
$^{150}$Department of Physics, Technion, Israel Institute of Technology, Haifa; Israel.\\
$^{151}$Raymond and Beverly Sackler School of Physics and Astronomy, Tel Aviv University, Tel Aviv; Israel.\\
% ...
$^{a}$ Also Affiliated with an institute covered by a cooperation agreement with CERN.\\
$^{b}$ Also at An-Najah National University, Nablus; Palestine.\\
$^{c}$ Also at APC, Universit\'e Paris Cit\'e, CNRS/IN2P3, Paris; France.\\
$^{d}$ Also at Borough of Manhattan Community College, City University of New York, New York NY; United States of America.\\
$^{e}$ Also at Center for High Energy Physics, Peking University; China.\\
$^{f}$ Also at Center for Interdisciplinary Research and Innovation (CIRI-AUTH), Thessaloniki; Greece.\\
% ...
$^{q}$ Also at Department of Physics, King's College London, London; United Kingdom.\\
% ...

\end{flushleft}
